# Generation of 10-dB squeezed light from a broadband waveguide optical parametric amplifier with improved phase locking method

Kazuki Hirota,[1,†] Takahiro Kashiwazaki,[2,†,*] Gyeongmin Ha,[1] Taichi Yamashima,[2] Pawaphat Jaturaphagorn,[1,3] Takumi Suzuki,[1] Kazuma Takahashi,[1] Akito Kawasaki,[1,4,5] Asuka Inoue,[2] Warit Asavanant,[4,5] Mamoru Endo,[1,4] Takeshi Umeki,[2] and Akira Furusawa[1,4,5**]

[1]*Department of Applied Physics, School of Engineering, The University of Tokyo, 7-3-1 Hongo, Bunkyo-ku, Tokyo 113-8656, Japan*
[2]*Device Technology Labs, NTT, Inc., 3-1 Morinosato Wakamiya, Atsugi, Kanagawa 243-0198, Japan*
[3]*Department of Physics, Faculty of Science, Kasetsart University, Bangkok 10900, Thailand*
[4]*Optical Quantum Computing Research Team, RIKEN Center for Quantum Computing, 2-1 Hirosawa, Wako, Saitama 351-0198, Japan*
[5]*OptQC Corp., 1-21-7 Nishi-Ikebukuro, Toshima-ku, Tokyo 171-0021, Japan*

[†]*These authors contributed equally.*
[*]*takahiro.kashiwazaki@ntt.com*
[**]*akiraf@ap.t.u-tokyo.ac.jp*

**Abstract:** We report generation of 10.1±0.2-dB squeezed light from a broadband periodically poled lithium niobate (PPLN) waveguide optical parametric amplifier (OPA). Based on our previous report where a similar PPLN waveguide shows 8.3-dB squeezing [T. Kashiwazaki *et al.*, Appl. Phys. Lett. **122**, 234003 (2023)], we reduce phase fluctuations and overall optical losses in the measurement system. In particular, we introduce a novel phase detection technique that does not require tapping a part of the squeezed light to get a phase locking signal. We use a phase-detection OPA seeded by a tapped probe and pump light before a squeezer OPA. This configuration breaks the conventional trade-off between generating a phase-locking signal with high-signal-to-noise ratio and suppressing degradation of squeezing level caused by optical tapping. With all these improvements, the phase fluctuation angle is reduced from 14 mrad to 9 mrad, and the total optical loss from 12% to 8%. Achieving more than 10 dB of squeezing by the broadband waveguide OPA is a significant step towards realization of fault-tolerant ultra-fast universal optical quantum computation.

## 1. Introduction

Squeezed light is one of the most essential resources for various quantum applications including quantum metrology [1-3], quantum communication [4,5], and quantum computation [6-9]. The performances of these applications are critically determined by two parameters: squeezing level and bandwidth of the squeezed light. A highly squeezed light has reduced quantum noise in one quadrature enabling high-precision measurements and low-noise quantum communication and computation, such as fault-tolerant quantum computation (FTQC) [10]. Furthermore, especially for continuous variable (CV) quantum computing, the broadband squeezed light enables quantum operation at extremely high clock rates and generation of large-scale entanglement through a time-domain-multiplexing (TDM) technique [8,9].

Since the first generation of squeezed light in 1985 [11], various approaches have been reported to improve the squeezing level with utilizing second- and third-order nonlinear optical effects [12]. Among them, optical parametric oscillators (OPOs), where a nonlinear crystal is placed inside an optical cavity, have long been the standard for achieving high squeezing levels. Using this method, squeezing levels as high as 15 dB have been reported around 3 megahertz (MHz) [13]. However, due to the resonant nature of optical cavities, the bandwidth of squeezed light generated by OPOs is inherently limited typically up to hundred MHz range [14]. That is, the broadband property of light, spanning the terahertz (THz) range, is not fully utilized due to the narrow bandwidth of OPOs.

To overcome this bandwidth limitation, cavity-free approaches using optical parametric amplifiers (OPAs) with nonlinear waveguides have attracted increasing attention. Waveguide OPAs show large nonlinear effect thanks to its high light-confinement in a small area with a long interaction propagation length. This strong nonlinearity enables single-pass optical parametric amplification, resulting in THz-order squeezing bandwidth. So far various improvements have been made to squeezed light generation using waveguide OPAs [15-23]. By using a periodically poled lithium niobate (PPLN) waveguide, 6-THz bandwidth of 6-dB squeezing [19] and 25-THz bandwidth of 4-dB [21] squeezing were succeeded in 2021 and in 2022, respectively. Thus, waveguide-based OPAs are attracting increasing attention because of their ability to generate broadband squeezed light. In 2024, continuous-wave 8.3-dB squeezed light was generated using a PPLN waveguide [23]. This squeezing level is minimum required noise level for FTQC using approximate Gottesman-Kitaev-Preskill (GKP) coding [24]. However, the achieved squeezing level remains insufficient for practical usage.

The reason for low squeezing level with a waveguide OPA is primarily attributed to intrinsic losses of the waveguide as well as imperfections of the detection system, including optical losses, electronic noises, and phase-locking fluctuations during measurement. In particular, accurate phase-stabilization becomes increasingly critical when measuring high-level squeezing. It is because achieving strong squeezing inevitably results in a large anti-squeezed quadrature. Consequently, even slight phase fluctuations can mix the anti-squeezed component into the squeezed quadrature, significantly degrading the measured squeezing level. For instance, achieving a squeezing level above 10 dB requires an anti-squeezing level above about 20 dB considering imperfection of detection system [13], resulting in a clearance between the two quadratures of more than 30 dB. In this case, a phase fluctuation of 10 mrad degrades the measured squeezing level, corresponding to an effective optical loss of approximately 1%. Typically, optical phase stabilization is performed based on feedback control. To enhance the phase-locking precision, not only the gain and bandwidth of the feedback must be optimized, but also the signal-to-noise ratio (SNR) of the phase error signal must be maximized. Conventionally, a probe light used for phase referencing is co-propagating with the pump into the squeezing OPA, and a small part of the output light is tapped after the OPA to generate the error signal [18, 23]. Here, increasing the tapping ratio to improve the SNR directly leads to additional optical loss of the squeezed light, creating a fundamental trade-off between phase precision and losses degrading squeezing.

In this paper, to overcome these limitations, we propose a novel phase-locking method that avoids tapping the squeezed light. Instead, a part of the probe and the pump light is extracted prior to the squeezer OPA and injected into a phase-detection OPA. The output light from the phase-detection OPA is used for the phase locking. Since no light is tapped after the squeezer OPA, the squeezing level is not degraded. Moreover, the power of the probe and pump used in the phase-detection OPA can be independently adjusted to generate a strong error signal with high SNR, thus enhancing the locking precision. As a result, we reduced the phase fluctuation angle from our previous condition of 14 mrad to 9 mrad and the optical losses in the system from 12% to 8% [23]. With these improvements, we succeeded in increasing the squeezing level from 8.3±0.2 dB to 10.1±0.2 dB by using a PPLN waveguide OPA.

## 2. Low-loss phase-detection method for phase locking

In the measurement of squeezed light, it is common to use a reference probe light for phase locking [18,23], since squeezed light itself does not possess a well-defined phase and amplitude like classical coherent light. The probe light is injected into an OPA along with the pump light that produces the squeezed light. A part of the output light is then tapped and converted into an electrical signal using a photodetector, as shown in Fig. 1(a). The probe light is typically phase-modulated in advance using a phase-modulator [18] or frequency-shifted using acousto-optic modulators (AOMs) [23]. The electrical signal obtained from a photodetector is demodulated at the modulation or shifted frequency, allowing the generation of an error signal. A control signal is generated by a proportional-integral-derivative (PID) servo controller to hold the error signal to zero, and this control signal is fed back to an actuator in the optical system to perform phase locking. In this configuration, the squeezed light itself is also partially tapped, resulting in the admixture of vacuum fluctuations and a consequent degradation of the squeezing level. While reducing the tapping ratio can mitigate this degradation, it simultaneously diminishes the strength of the signal used for phase locking. As a result, the SNR deteriorates, leading to reduced precision in phase stabilization. It also degrades the squeezing level.

To solve this trade-off between squeezing degradation and phase-locking accuracy, we propose a novel phase-locking method, in which the pump and probe are split prior to the squeezer OPA and launched into another OPA for phase detection, as illustrated in Fig. 1(b). In this method, phase locking can be achieved if there is no medium that significantly disturbs the relative phase between the pump and the probe light in the optical path after separating the pump and probe light and before entering the phase-detection OPA. In practice, since these light travel along nearly identical optical paths, their relative phase is well preserved. In this configuration, it is no longer necessary to tap the generated squeezed light, thereby avoiding degradation of the squeezing level due to vacuum noise contamination. Furthermore, the intensities of the pump and probe light injected into the phase-detection OPA can be independently adjusted, allowing the generation of an error signal with a high SNR. In conventional methods, the probe beam intensity must be kept relatively low, typically in the order of a few microwatts, for suppressing the amount of probe light entering the homodyne detector. In contrast, the novel method allows the probe light to be as strong as a few milliwatts, which significantly improves the SNR.

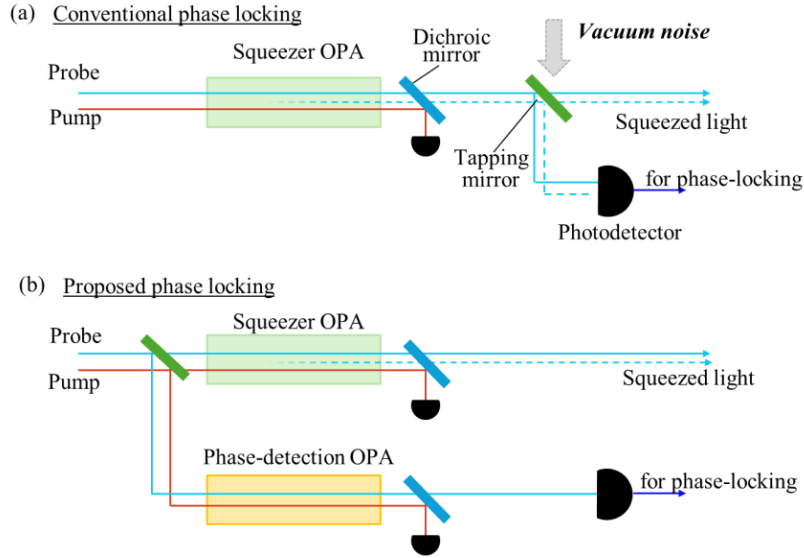

Fig. 1. Phase locking method. (a) Conventional phase locking with tapping a part of the squeezed light after a squeezer OPA. (b) Phase locking method without tapping of squeezed light. Both probe and pump light are tapped before the squeezer OPA and launched into another OPA for phase detection. Inside the phase-detection OPA, optical parametric process occurs according to the relative optical phase difference between probe and pump.

## 3. Experiment

### 3.1 Experimental setup for squeezing level measurement

Figure 2 shows squeezing level measurement setup with the novel phase-locking method. The squeezed light is generated by PPLN waveguide OPA, which is fabricated by the same method in Ref [19]. Its waveguide length is 45 mm, and waveguide core has near rectangle shape with 8-µm width and 8-µm height. A seed laser system (NKT photonics, Koheras harmonik) outputs both fundamental and second harmonic light, whose wavelengths are 1545.32 nm and 772.66 nm. The fundamental light is separated for local oscillator (LO) and probe light for phase locking of a squeezer OPA. Both paths for LO and probe have an electro-optic phase modulator (iXblue, MPX-LN-0.1) (EOM) for phase locking. In the probe path, we use two acoustic-optic modulators (Chongquing Smart Science &Technology Development, SGTF-40-1550-1P) (AOMs) for shifting the frequency of probe light. The shift frequency is 1 MHz. The probe light and second harmonic pump light are combined by a dichroic mirror and separated into two paths for the squeezer OPA and another OPA for phase-detection. The pump and probe light is coupled into the waveguide OPA by a lens with focal length of 4.5 mm. In the phase-detection OPA, the probe light is amplified by nonlinear optical parametric process and generates phase locking signal. After the signal detection by an InGaAs photodiode, the error signal is sent for a hand-made phase-locking PID servo controlling system. Thanks to this configuration, we can lock the relative optical phase between pump and probe light inside the squeezer OPA without tapping a part of generated squeezed light. The generated squeezed light is measured by balanced homodyne detection with handmade balanced detector, which has two InGaAs photodiodes (Laser Components, IGHQEX0500-1550-10-1.0-SPAR-TH-40). The quantum efficiencies of these photodiodes are both about 99%. The signal from the detector is analyzed by an electrical spectrum analyzer (Keysight, N9010B) (ESA). A filter cavity is employed in the optical path for LO to suppress the ASE noise accompanying the LO light. This is critically important for the measurement of broadband squeezed light. It is because the beat between the broadband squeezed light and the broadband ASE light appears as additional noise in balanced homodyne detection. In this experiment, we improve several optical components to suppress optical losses from previous work [23].

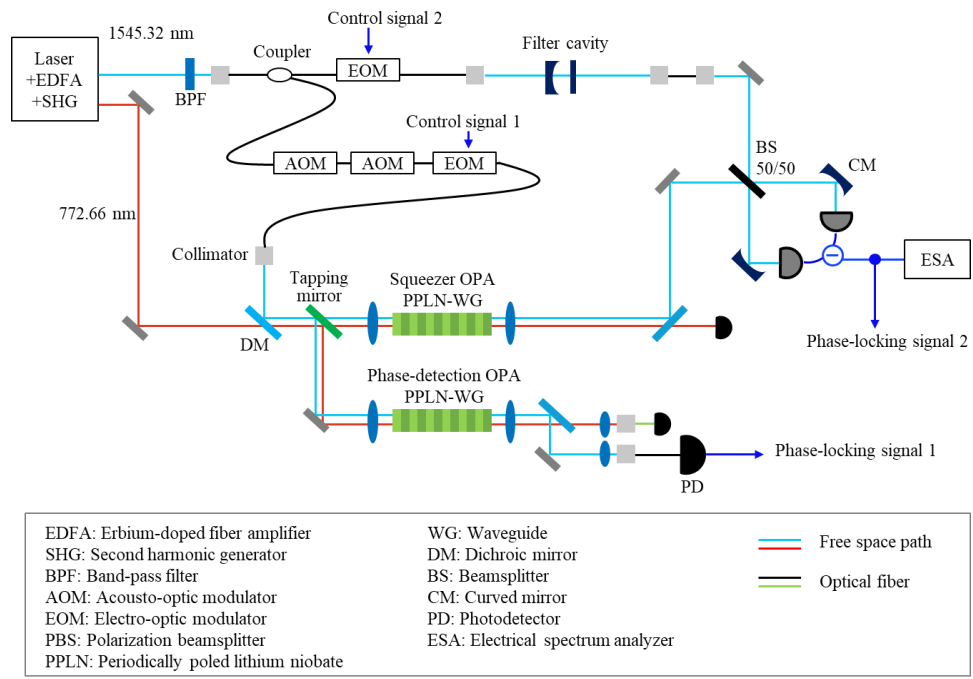

Fig. 2. Experimental setup for measuring noise of squeezed vacuum state of light from a PPLN waveguide. To suppress the effect of ASE light, we insert a filter cavity in an optical path for LO. Furthermore, we use another PPLN to generate phase locking signal.

## 3.2 Results

Figure 3 shows squeezing levels detected by the spectrum analyzer in zero span mode. The measurement center frequency is 3 MHz, a resolution bandwidth (RBW) is 1 MHz, and a video bandwidth (VBW) is 100 Hz. The output pump power immediately after the squeezer OPA is estimated to be 640 mW. The squeezing level is measured as 10.1±0.2 dB without any loss correction and circuit-noise correction. Here, the standard deviation of the phase-locked squeezed noise level is 0.07 dB. Assuming a Gaussian distribution, the measurement error, defined as the range containing 99.7% of the observed values, can be estimated to be approximately ±0.2 dB. The measured anti-squeezing level is 20.2±0.2 dB. In Fig. 4, the measured noise of squeezed light with phase-scanned LO is also shown.

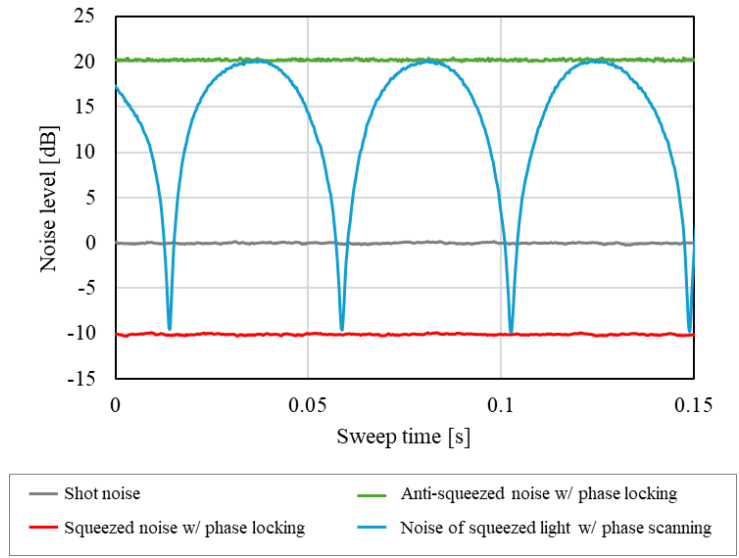

Fig. 3. Measured noise levels by homodyne detection. The gray line is shot noise level. The red and green lines are anti-squeezed and squeezed noise levels with phase locking. The blue line is the noise for squeezed light with phase scanning. The center frequency is 3 MHz, RBW is 1 MHz, and video bandwidth is 100 Hz.

Figure 4(a) shows frequency spectra of the noise at a pump power of 640 mW. The RBW and the VBW in this measurement are 1 MHz and 100 Hz, respectively. The peak at 1 MHz corresponds to the shift frequency of probe light for phase locking. Thanks to the broad bandwidth of the squeezed light from waveguide OPA, the squeezing level is kept for high frequency even with the circuit noise increment. Fig. 4(b) shows normalized noise levels with subtracting the effect of the circuit noise. At the measurement frequency of 100 MHz, the squeezing level is over 8 dB. Slight degradation observed at high frequencies can be attributed to the deterioration in the detection efficiency of the photodiodes, whose bandwidth is 35 MHz.

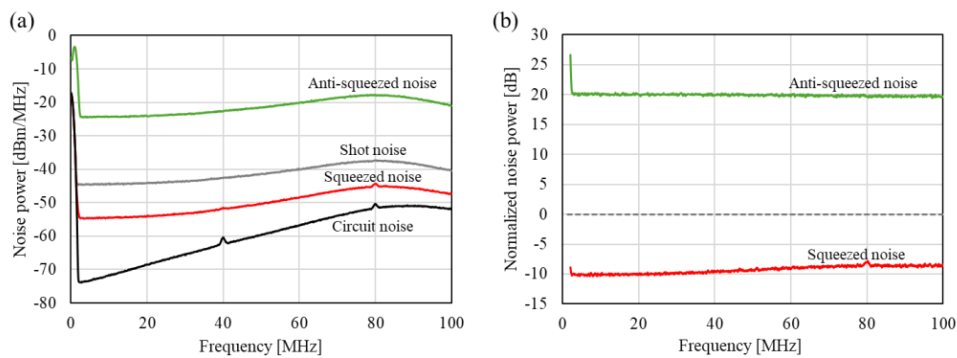

Fig. 4. (a) Measured noise levels for squeezed light as a function of frequency. RBW is 1 MHz. VBW is 100 Hz. (b) Normalized squeezed noise level with subtracting circuit noise.

## 4. Discussions

### 4.1 Phase noise

To investigate the phase locking accuracy, we plot squeezing levels as a function of the pump power after the squeezer OPA, as shown in Fig. 5. Theoretical curve is also drawn with various phase locking fluctuations with the following equations [25-27];

$$R'_{\pm}(\tilde{\theta}) \approx R_{\pm} \cos^2 \tilde{\theta} + R_{\mp} \sin^2 \tilde{\theta}.$$

Here $\tilde{\theta}$ is a standard deviation of relative phase angle difference between the LO and anti-squeezed or squeezed quadrature with assuming small angle. The fluctuation is assumed as the type of gaussian fluctuation. $R_+$ and $R_-$ are actual anti-squeezed and squeezed noise levels, which are described in following equation.

$$R_{\pm} = L + (1 - L) \cdot \exp(\pm 2\sqrt{\alpha P}).$$

Here, $\alpha$ is second-harmonic-generation (SHG) efficiency of the waveguide, $P$ is pump power, $L$ is effective optical loss for squeezed light. By using least square law fitting, we evaluate the fluctuation angle, optical loss, and SHG efficiency in this experiment as 9 mrad, 8%, and 906 %/W, respectively. In Fig. 5, a theoretical curve is also shown with parameters of 14-mrad fluctuation angle and 12% optical loss, which are the parameters of our previous condition [23]. It is obvious that the squeezing levels have improved from previous work. For more precise phase locking, various noises should be reduced such as the noise of the light emitted from the seed laser by inserting a high-finesse optical cavity.

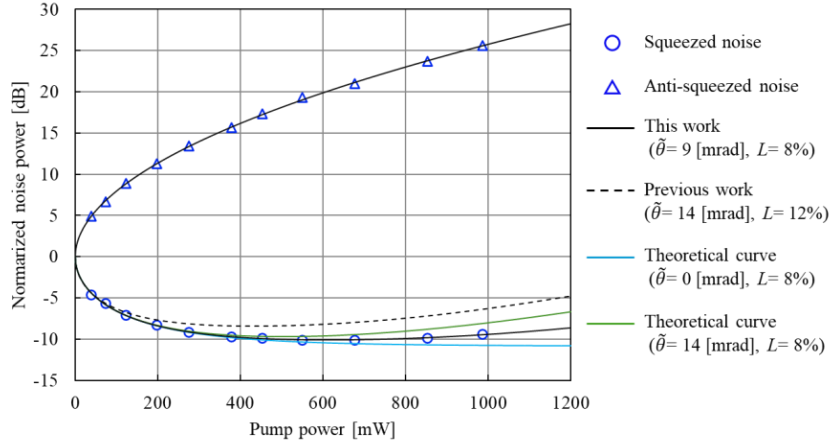

Fig. 5. Measured noise levels for squeezed light with various pump powers. Solid black line is a fitted theoretical curve with fitting parameters of a pump power, a second-harmonic efficiency, an optical loss, and a phase fluctuation. Dashed line is drawn with $\tilde{\theta}$ as 14 mrad and $L$ as 12%, which are the parameters same as the previous research showing 8.3 dB squeezing. Blue line is drawn with $\tilde{\theta}$ as 0 rad and $L$ as 8%. Green line is drawn with $\tilde{\theta}$ as 14 mrad and $L$ as 8%

### 4.2 Optical loss

In this section, we analyze optical loss in the experiment with 640 mW of pump power. The measured squeezing and anti-squeezing levels are $10.1 \pm 0.2$ dB and $20.2 \pm 0.2$ dB, respectively. As discussed above, the phase fluctuation and the optical loss are estimated to be 9 mrad and 8%. This optical loss can be divided into four parts: a waveguide loss, the optical loss in the measurement setup, equivalent loss due to electrical noise, and the loss due to mode mismatch between the squeezed light and LO.

First, the waveguide loss is determined by the propagation loss per unit length and the nonlinear efficiency of the waveguide, as described in Ref [28]. This is because the squeezed light is generated while simultaneously undergoing continuous optical attenuation in the waveguide. The propagation loss of the PPLN waveguide used in this experiment is approximately 0.1 dB/cm [19]. Consequently, the effective loss in the waveguide is estimated to be 2% with the pump power of 640 mW.

Second, the optical loss in the measurement setup consists of the loss of the optical components and photo-electro conversion loss at photodiodes. The loss of optical components, such as mirrors and lenses, is measured as 1%. The photo-electro conversion loss is estimated to be about 1%.

Third, to evaluate the equivalent loss due to the circuit noise, we measured the intensities of shot noise and circuit noise. The clearance between the circuit noise and the shot noise is about 28 dB at a frequency around 3 MHz with 15.4 mW of LO power. Therefore, the equivalent loss due to circuit noise is lower than 0.2%.

Finally, we estimate the loss due to the mode mismatch to be about 4% by subtracting the previously mentioned losses from the total optical loss of 8% obtained by theoretical fitting in Fig. 5. Although it is difficult to directly measure the interference efficiency between the squeezed light and the LO light, we consider the estimated value to be reasonable. This is because the maximum interference efficiency between the probe light, which is amplified by the squeezer OPA, and the LO light was approximately 99% resulting in about 2% of optical loss. The origin of the 2% discrepancy could not be identified in this experiment; however, it is considered to a result of a slight difference of the spatial modes between the squeezed light and the probe light. It indicates that employing a spatial phase modulator in the LO path to improve the mode overlap [22] could potentially lead to a higher level of squeezing. Furthermore, there are several other improvements that can be considered to achieve higher squeezing levels as future works. For example, one possible approach is to enhance the photodiode detection efficiency by re-injecting the light reflected from the photodiode into it [13]. In addition to reducing the intrinsic loss of the waveguide through fabrication process improvements, the core size of the waveguide can also be reduced to increase the nonlinear efficiency per unit length. However, a smaller core may lead to lower coupling efficiency and higher propagation loss, so an optimal trade-off must be carefully evaluated.

## 5.  Conclusion

We report generation of a $10.1 \pm 0.2$-dB squeezed light from a broadband PPLN waveguide by using low-loss phase locking method to reduce phase-locking fluctuation and optical loss of the measurement system. We use a phase-detection OPA seeded by tapped probe and pump light before a squeezer OPA to generate a phase locking signal. This configuration breaks the conventional trade-off between phase noise and optical loss by enabling precise phase locking without deteriorating squeezed light. As a result, we successfully reduced the phase fluctuation angle and optical loss to 9 mrad and 8% from our previous experiment [23]. Achieving more than 10 dB of squeezing can relax the requirement for generating approximate GKP states [24]. Thus, it is a significant step towards realization of practical fault-tolerant ultra-fast universal optical quantum computation. In addition, this result also contributes to the advancement of recently launched analog quantum computing platforms [29] having potential for applications in artificial intelligence, particularly in the development of optical neural networks.

**Funding.** Japan Science and Technology Agency (JPMJMS2064, JPMJPR2254). Japan Society for the Promotion of Science (23K13040). National Research Council of Thailand (N41A640473).

**Acknowledgment.** The authors acknowledge supports from UTokyo Foundation and donations from Nichia Corporation of Japan. T.S. and A.K acknowledge financial support from The Forefront Physics and Mathematics Program to Drive Transformation (FoPM), a World-leading Innovative Graduate Study (WINGS) Program, the University of Tokyo. A.K. acknowledges financial support from Leadership Development Program for Ph.D (LDPP), the University of Tokyo. M.E. and W.A. acknowledge support from Research Foundation for OptoScience and Technology. A part of this work was performed for Council for Science, Technology and Innovation (CSTI), Cross-ministerial Strategic Innovation Promotion Program (SIP), "Promoting Application of Advanced Quantum Technologies to Social Challenges" (Project management agency: QST).

**Disclosures.** The authors declare no competing financial interest.

**Data availability.** The data that support the findings of this work are available from the corresponding author upon reasonable request.

## References

1. C. M. Caves, "Quantum-mechanical noise in an interferometer," Phys. Rev. D 23, 1693 (1981).
2. V. Giovannetti, S. Lloyd, and L. Maccone, "Quantum-Enhanced Measurements: Beating the Standard Quantum Limit," Science 306, 1330 (2004).
3. H. Grote, K. Danzmann, K. L. Dooley, R. Schnabel, J. Slutsky, and H. Vahlbruch, "First Long-Term Application of Squeezed States of Light in a Gravitational-Wave Observatory," Phys. Rev. Lett. 110, 181101 (2013).
4. N. J. Cerf, M. Levy, and G. V. Assche, "Quantum distribution of gaussian keys using squeezed states," Phys. Rev. A 63, 052311 (2001).
5. L. S. Madsen, V. C. Usenko, M. Lassen, R. Filip, and U. L. Andersen, "Continuous variable quantum key distribution with modulated entangled states," Nature Communications 3, 1083 (2012).
6. S. L. Braunstein and P. van Loock, "Quantum information with continuous variables," Rev. Mod. Phys. 77, 513 (2005).
7. H. Aghaee Rad, T. Ainsworth, R. N. Alexander et al, "Scaling and networking a modular photonic quantum computer," Nature 638, 912 (2025).
8. W. Asavanant and A. Furusawa, "Optical Quantum Computers a Route to Practical Continuous Variable Quantum Information Processing," AIP Publishing, Melville, New York (2022).
9. S. Takeda and A. Furusawa, "Toward large-scale fault-tolerant universal photonic quantum computing," APL Photonics 4, 060902 (2019).
10. D. Gottesman, A. Kitaev, and J. Preskill, "Encoding a qubit in an oscillator," Phys. Rev. A 64, 012310 (2001).
11. R. E. Slusher, L. W. Hollberg, B. Yurke, J. C. Mertz, and J. F. Valley, "Observation of squeezed states generated by four-wave mixing in an optical cavity," Phys. Rev. Lett. 55, 2409 (1985).
12. U. L. Andersen, T. Gehring, C. Marquardt, and G. Leuchs, "30 years of squeezed light generation," Phys. Scr. 91, 053001 (2016).
13. H. Vahlbruch, M. Mehmet, K. Danzmann, and R. Schnabel, "Detection of 15 dB squeezed states of light and their application for the absolute calibration of photoelectric quantum efficiency," Phys. Rev. Lett. 117, 110801 (2016).
14. N. Takanashi, W. Inokuchi, T. Serikawa, and A. Furusawa, "Generation and measurement of a squeezed vacuum up to 100 MHz at 1550 nm with a semi-monolithic optical parametric oscillator designed towards direct coupling with waveguide modules," Opt. Express 27, 18900 (2019).
15. M. E. Anderson, J. D. Bierlein, M. Beck, and M. G. Raymer, "Quadrature squeezing with ultrashort pulses in nonlinear-optical waveguides," Opt. Lett. 20, 620 (1995).
16. D. K. Serkland, M. M. Fejer, R. L. Byer, and Y. Yamamoto, "Squeezing in a quasi-phase-matched LiNbO$_3$ waveguide," Opt. Lett. 20, 1649 (1995).
17. K. Yoshino, T. Aoki, and A. Furusawa, "Generation of continuous-wave broadband entangled beams using periodically poled lithium niobate waveguides," Appl. Phys. Lett. 90, 041111 (2007).
18. T. Kashiwazaki, N. Takanashi, T. Yamashima, T. Kazama, K. Enbutsu, R. Kasahara, T. Umeki, and A. Furusawa, "Continuous-wave 6-dB-squeezed light with 2.5-THz-bandwidth from single-mode PPLN waveguide," APL Photonics 5, 036104 (2020).
19. T. Kashiwazaki, T. Yamashima, N. Takanashi, A. Inoue, T. Umeki, and A. Furusawa, "Fabrication of low-loss quasi-single-mode PPLN waveguide and its application to a modularized broadband high-level squeezer." Applied Physics Letters 119.25 (2021).

20. Y. Taguchi, K. Oguchi, Z. Xu, D. Cheon, S. Takahashi, Y. Sano, F. Harashima, and Y. Ozeki, "Phase locking of squeezed vacuum generated by a single-pass optical parametric amplifier." Optics Express 30, 8002 (2022).
21. R. Nehra, R. Sekine, L. Ledezma, Q. Guo, R. M. Gray, A. Roy, and A. Marandi, "Few-cycle vacuum squeezing in nanophotonics," Science 377, 6612 (2022).
22. A. Jorge, J. Takai, and T. Hirano. "Highly efficient measurement of optical quadrature squeezing using a spatial light modulator controlled by machine learning." Optics Continuum 2, 933 (2023).
23. T. Kashiwazaki, T. Yamashima, K. Enbutsu, T. Kazama, A. Inoue, K. Fukui, M. Endo, T. Umeki, and A. Furusawa "Over-8-dB squeezed light generation by a broadband waveguide optical parametric amplifier toward fault-tolerant ultra-fast quantum computers." Appl. Phys. Lett. 122, 23 (2023).
24. K. Fukui, "High-threshold fault-tolerant quantum computation with the Gottesman-Kitaev-Preskill qubit under noise in an optical setup," Phys. Rev. A 107, 052414 (2023).
25. S. Suzuki, H. Yonezawa, F. Kannari, M. Sasaki, and A. Furusawa, "7 dB quadrature squeezing at 860 nm with periodically poled KTiOPO4," Appl. Phys. Lett. 89, 061116 (2006).
26. T. C. Zhang, K. W. Goh, C. W. Chou, P. Lodahl, and H. J. Kimble, "Quantum teleportation of light beams," Phys. Rev. A 67, 033802 (2003).
27. Aoki, G. Takahashi, and A. Furusawa, "Squeezing at 946 nm with periodically poled KTiOPO4," Opt. Express 14, 6930 (2006).
28. T. Yamashima, T. Kashiwazaki, T. Suzuki, R. Nehra, T. Nakamura, A. Inoue, T. Umeki, K. Takase, W. Asavanant, M. Endo, and A. Furusawa, "All-optical measurement-device-free feedforward enabling ultra-fast quantum information processing," Opt. Express 33, 5769 (2025).
29. S. Yokoyama, A. Sakaguchi, W. Asavanant, K. Takase, Y. Chen, H. Nagayoshi, J. Yoshikawa, T. Kashiwazaki, A. Inoue, T. Umeki, T. Hashimoto, T. Hiraoka, A. Furusawa, and H. Yonezawa, "Full-stack Analog Optical Quantum Computer with A Hundred Inputs," arXiv:2506.16147.